\documentclass[useAMS]{mn2e}
\usepackage{times}
\usepackage{amssymb}
\usepackage{rotating}
\input{epsf}

\title[The soft X-ray excess in AGN]
{Intrinsic disc emission and the Soft X-ray Excess in AGN}

\author[C. Done, S. W. Davis, C. Jin, O. Blaes, M. Ward ]
{Chris Done$\thanks{E-mail:chris.done@durham.ac.uk}^1$, S. W. Davis$^2$,
  C. Jin$^1$, O. Blaes$^3$, M. Ward$^1$
\\
$^1$ Department of Physics, University of Durham, South Road,
Durham DH1 3LE, UK\\
$^2$ CITA, University of Toronto, 60 George Street, Toronto, ON, M5S
3H8, Canada\\
$^3$ Department of Physics, Broida Hall, UCSB, CA 93106-9530, USA\\
}

\date{Submitted to MNRAS}
\pagerange{\pageref{firstpage}--\pageref{lastpage}} \pubyear{2009}

\def\msun{M$_\odot$\ }

\begin{document}

\topmargin = -0.5cm

\maketitle

\label{firstpage}

\begin{abstract}

Narrow Line
Seyfert 1 (NLS1) galaxies have low mass black holes and
mass accretion rates close to (or exceeding) Eddington, so a
standard blackbody accretion disc should peak in the EUV. However,
the lack of true absorption opacity in the disc means that 
the emission is better approximated by a colour temperature 
corrected blackbody, and this colour temperature correction is
large enough ($\sim 2.4$) that the bare disc emission from a zero
spin black hole can extend
into the soft X-ray bandpass. Part of the soft X-ray excess seen in
these objects must be intrinsic emission from the disc unless the
vertical structure is very different to that predicted. 

Nontheless, this is not the whole story even for the extreme NLS1 as
the shape of the soft excess is much broader than predicted by a bare
disc spectrum, indicating some Compton upscattering by warm, optically
thick material.  We associate this with the disc itself, so it must
ultimately be powered by mass accretion.  We
build an energetically self consistent model assuming that the
emission thermalises to a (colour temperature corrected) blackbody
only at large radii. At smaller radii the gravitational energy
is split between powering
optically thick Comptonised disc emission (forming the soft X-ray
excess) and an optically thin corona above the disc (forming the tail
to higher energies). We show examples of this model fit to the extreme
NLS1 REJ1034+396, and to the much lower Eddington fraction Broad Line
Seyfert 1 PG1048+231. We use these to guide our
fits and interpretations of three template spectra made
from co-adding multiple sources to track out a sequence of AGN spectra
as a function of $L/L_{Edd}$. 

Both the individual objects and template spectra show the surprising
result that the Compton upscattered soft X-ray excess decreases in
importance with increasing $L/L_{Edd}$. The strongest soft excesses
are associated with low mass accretion rate AGN rather than being tied
to some change in disc structure around Eddington. We argue that this 
suggests a true break in accretion flow properties between
stellar and supermassive black holes.

The new model is publically available within the {\sc xspec} spectral 
fitting package. 

\end{abstract}


%==============================================
\section{Introduction} \label{sec:introduction}
%==============================================

High mass accretion rate AGN ubiquitously show a 'soft X-ray excess',
where the X-ray data below 1~keV lie above the low energy
extrapolation of the best fitting 2-10~keV power law. The origin of
this feature is currently the subject of active research, but plainly
it is generally too hot to be the standard optically thick disc.
Relativistic, optically thick, geometrically thin accretion disc
models (Shakura \& Sunyaev 1973; Novikov \& Thorne 1973) give a
maximum effective temperature of the accreting material of $kT\sim 10
(\dot{m}/M_8)^{1/4}$ ~eV, where $\dot{m}$ is mass accretion rate in
units of Eddington (so that
$\dot{m}=\dot{M}/\dot{M}_{Edd}=L/L_{Edd}$), and
$M_8=M/10^8$~M$_\odot$. Thus the disc spectrum of a typical quasar with
a $10^8$\msun black hole accreting at $L/L_{Edd}=0.2$ peaks in the EUV
range, at $\sim 2.7k_B T \sim 20$~eV, and the Wein shape above this
energy means that it makes very little contribution to the soft X-ray
flux above 0.3~keV (e.g Laor et al. 1997).

This assumes that the emission is able to thermalise completely. This
is unlikely to always be true as AGN discs can be dominated by
electron scattering rather than absorption in regions where the
temperature is well above the Hydrogen ionisation energy of 13.7~eV or
$10^5$~K (e.g. Ross, Fabian \& Mineshige 1992). Such high disc
temperatures are found only for a combination of lower black hole mass
and higher $L/L_{Edd}$, so this is an effect which will be
preferentially important in the Narrow Line Seyfert 1 subclass of AGN,
objects which often have particularly prominant soft X-ray excesses
(see e.g. Boller, Brandt \& Fink 1996).

While this could increase the extent of  the disc emission in the 
soft X-ray regime, its predicted shape is a 
Wien tail which drops much more sharply with energy than the observed,
more gradual decline of the soft X-ray excess (Bechtold et al. 1987;
Laor et al. 1997). Instead, the observed spectral shape can be fit by
Compton upscattering.  Clearly there is at least one Compton
upscattering 
region required to produce the power law seen beyond 2~keV, above
which the soft excess is measured. The high energy extent of this
power law means that the electrons are hot ($kT_e\sim 100$~keV), and
have low optical depth, $\tau\sim 1$ (e.g Zdziarski et al. 1995; Zdziarski, Johnson \& Magdziarz 1996). To
additionally make the very different shape of the soft excess from
Compton upscattering requires that there is another electron
population, one which has much lower temperature and higher optical
depth (Bechtold et al. 1987, Czerny \& Elvis 1987).

Physically, this component could arise in a transition region between
the disc and hot corona, either separated radially (Magdziarz et
al.~1998) or vertically (Janiuk, Czerny \& Madejski 2001).  However, a
key problem with this interpretation is that the temperature
associated with the Comptonisation region is observed to remain
remarkably constant at $\sim 0.2$~keV in all high mass accretion rate AGN
(Gierlinksi \& Done 2004; Czerny et al. 2003). A
Comptonisation region with optical depth $\tau$, heated by power
$\ell_h$ and cooled by illumination of soft seed photon power $\ell_s$
has temperature which is determined by both $\ell_h/\ell_s$ {\em
and} $\tau$ (see e.g. Done 2010 section 1.3.2). A fairly constant
temperature then requires that these two parameters are
correlated, which seems finetuned. 
Indeed, in black hole binaries where the coronal spectra sometimes
require two Comptonisation components, the temperature of the
optically thicker/lower temperature component clearly varies (see e.g.
Fig 4 of  Kubota \& Done 2004). 

\begin{figure*}
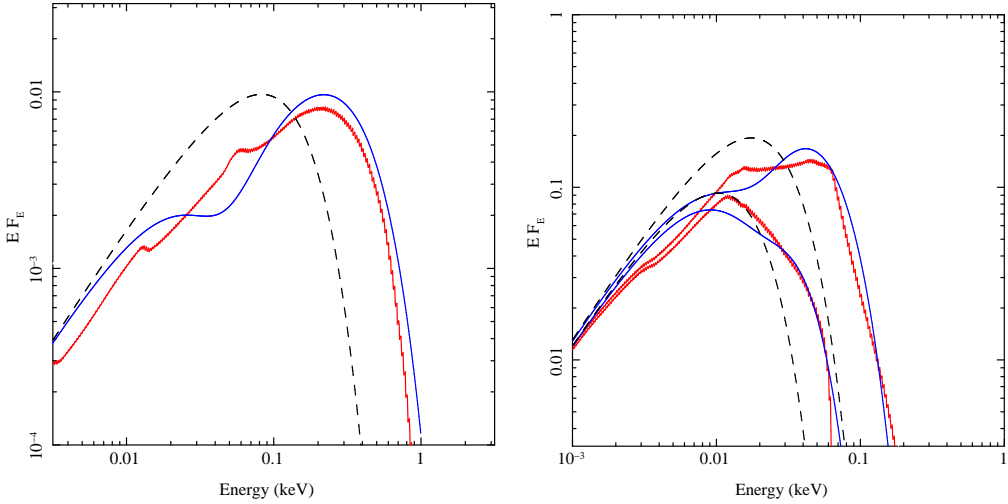

\centering
\begin{tabular}{l|r}
\leavevmode  
\epsfxsize=6.5cm \epsfbox{optxagn_hubeny3.ps} &
\epsfxsize=6.5cm \epsfbox{optxagn_hubeny45_3.ps}\\
\end{tabular}
\caption{Accretion disc spectra for a Schwarzschild black hole.  The
black dashed line is the spectrum from a standard Shakura-Sunyaev disc
assuming that the energy thermalises completely (black dashed line),
while the blue solid line has a single colour temperature correction
applied to all radii with $T>T_{scatt}$ and the red line is the disc 
spectrum calculated from the full radiative transfer code.  a) has $10^6$\msun and
$L/L_{Edd}=1$ with $f_{col}=2.6$ for $T>T_{scatt}=10^5$~K while b) has
$10^8$\msun, $L/L_{Edd}=0.2$ with $f_{col}=2$ together with $2\times
10^8$\msun and $L/L_{Edd}=0.05$ and $f_{col}=1.8$, both with
$T_{scatt}=4\times 10^4$~K. 
}
\label{fig:optxhub}
\end{figure*}

An alternative scenario is where the low temperature, optically thick
Comptonisation is instead the intrinsic spectrum of the disc
itself. Substantial changes in disc structure from that predicted by a
Shakura-Sunyaev disc might be expected as the accretion rate
approaches the Eddington limit.  This is especially attractive for
some extreme narrow line Seyfert 1 objects, where the luminosity is
around Eddington and where the soft excess is observed to be less
variable than the 2-10~keV power law (Middleton et al. 2009; Jin et al
2009). Comptonisation of the disc emission could occur in the disc
itself if there is enhanced dissipation in the upper layers of the
disc, over and above that expected from the Shakura-Sunyaev
prescription (e.g. Davis et al 2005). While this could be produced by
the Magneto-Rotational Instability (MRI: the physical mechanism
transporting angular momentum: Balbus \& Hawley 1991) there is
currently no straightforward way to parameterise the effects of this
with increasing mass accretion rate (e.g. Blaes et al 2011).  However,
it is clear that as the mass accretion rate approaches Eddington there
are several other effects which could act to distort the disc
spectrum, such as if the MRI drives large scale turbulence (Socrates,
Davis \& Blaes 2004) or if radially advected radiation associated with
a slim disc (Abramowicz et al 1989) can be released in the plunging
region (Sadowski  2009).

Thus there are some possibilities to produce the soft excess from the
disc itself (as opposed to a separate corona), where the inner disc
emission emerges as Compton upscattered flux rather than as a colour
temperature corrected blackbody. If so, then a key aspect of the
resulting spectrum is that it is powered by the mass accreting through
the disc, so the total energy is set by the black hole mass, spin and
mass accretion rate through the outer disc. The former can now be
constrained via the mass scaling relationships (e.g. Kaspi et al
2000).  The level of optical/UV flux from the disc then
directly determines the mass accreting through the outer disc 
(e.g. Davis \& Laor 2011). Here
we build a model incorporating these ideas, and demonstrate that at
least part of the soft excess seen in the lowest mass, highest
mass accretion rate AGN must be intrinsic emission from the
disc itself.

\section{Disc models}

We build a standard blackbody disc model, for a mass accretion rate
$\dot{M}$~g/s onto a black hole of mass $M$ and spin $a_*$ using the
effective temperature as a function of radius as derived from the
Novikov-Thorne relativistic disc emissivity. We integrate the fluxes
from each annulus in the disc over all radii from $r_{out}$ down to
the last stable orbit and assume that the emission is isotropic.
The black dashed line in Fig. \ref{fig:optxhub}a shows the spectrum produced
by this for a Schwarzchild black hole of mass
$M=10^6$\msun, accreting at the Eddington limit, with
$r_{out}=10^5R_g$ where $R_g=GM/c^2$. This has a peak temperature of
$4\times 10^5$~K, which is too cool to 
extend to soft X-ray energies. Extreme spin and/or
substantial Compton upscattering is required in order to make 
significant soft X-ray emission from a disc where the emission 
thermalises to the local blackbody temperature (e.g. Laor et al. 1997).

However, disc spectra is
more complex than a simple sum of blackbodies at
the effective temperature, $B_\nu(T)$, as the true absorption opacity,
$\kappa_{abs}(\nu)$ can be substantially smaller than the electron
scattering opacity, $\kappa_T$, leading to a modified blackbody
spectrum, $I_\nu(T)\sim \sqrt{\kappa_{abs}/\kappa_{tot}}B_\nu (T)$, where
$\kappa_{tot}= \kappa_{abs} + \kappa_T$ (Shakura \& Sunyaev 1973;
Czerny \& Elvis 1987). 

Absorption opacity, $\kappa_{abs}$, depends on temperature, density
and frequency whereas the electron scattering opacity is
constant. Thus $\kappa_{abs}/\kappa_{es}$ and hence the detailed
spectrum is sensistive to the vertical structure of the disc.  For a
given density, the absorption opacity is lower at higher temperature,
so the effective photosphere where $\tau_{eff} \approx \tau_T
\sqrt{\kappa_{abs}(\nu)/\kappa_T}= 1$ is at a greater depth in the
disc. The temperature increases with depth, so the difference between
this temperature and the surface temperature increases, giving an
increased colour temperature correction.  However, the increased
depth of formation also increases the number of scatterings before
escape in the lower temperature material between the photosphere and
the surface, increasing Compton downscattering which lowers the colour
temperature correction. Davis et al (2006) shows that these two
competing effects lead to the effective saturation of the colour
temperature correction to a value 
$$f_{col}\sim (72/T_{keV})^{1/9}\eqno{(1)}$$ (their Equation
A13). They derived this in the context of black hole binary (BHB)
discs, where the typical effective temperature of $\sim 1$~keV gives a
predicted $f_{col}\sim 1.6$ which has very little dependence on mass
accretion rate as $T\propto (L/L_{Edd})^{1/4}$ so $f_{col}\propto
(L/L_{Edd})^{1/36}$. This is consistent with the observed $L\propto
T^4$ behaviour (which implies constant colour temperature correction
as well as constant inner radius) seen in BHB disc dominated spectra
over a wide range in $L/L_{Edd}$ (Kubota et al 2001; Gierlinski \&
Done 2004).

While Equation 1 was derived for BHB discs, its assumptions should
also hold in AGN discs. The first, and most important assumption is
that electron scattering dominates, i.e. that $\kappa_{abs}\ll
\kappa_{es}$. Typically, $\kappa_{abs}\propto nT^{-\beta}$, where $n$
is the density, and $\beta=3.5$ for free-free absorption. However,
bound-free absorption edges are typically more important.  Each
individual edge gives $\beta=2.7-3$ but added together the effect from
all ionisation states of all elements coincidentally gives $\beta\sim
3.5$. BHB discs are at much higher temperature than AGN discs as
$T\propto M^{-1/4}$, typically with $T\sim 10^7$~K where the
absorption opacity is dominated by the metal edges. These have much
lower abundance than Hydrogen and Helium, which typically dominate the
bound-free opacity at AGN disc temperatures of $T\sim 10^5$~K.
Nonetheless, this increase in opacity with decreasing temperature is
more than offset by the lower density of AGN discs, as $n\propto
M^{-1}$ so $\kappa_{abs}\propto M^{-1} [M^{-1/4}]^{-3.5}\propto
M^{-1/8}$. Thus for the same luminosity in terms of Eddington,
$L/L_{Edd}$, scattering is more important in an AGN disc than in a
black hole binary (BHB) disc.

The other assumptions on the temperature and density structure of the
outer layers of the radiation pressure dominated disc should be
equally applicable to AGN and BHB discs. The only other caveat is that
there is no significant additional dissipation of energy above the
effective photosphere i.e. we must be able to treat it as an
atmosphere, otherwise the colour temperature correction increases
further (Done \& Davis 2008).  Hence we should be able to apply
Equation 1 to AGN discs also, and we can test this by comparison with
full disc radiative transfer models.  Pioneering calculations by Ross,
Fabian \& Mineshige (1992) showed $f_{col}\sim 2.4$ for a $10^6$\msun
Schwarzchild black hole accreting at $L/L_{Edd}=0.33$. These disc
parameters predict a peak disc temperature of $3\times 10^5$~K, so
Equation 1 gives $f_{col}=2.4$, in excellent agreement with Ross et al
(1992).

We use the Hubeny et al (2001) models, recalculated as in Davis \&
Hubeny (2006), to independently test Equation 1. This uses the
Novikov-Thorne emissivity, but includes full radiative transfer
through the vertical structure of the disc at each radius, taking into
account both Compton scattering and full metal opacities (bound-free).
Unlike our model, the resulting radiation is then propagated to the
observer with full General Relativistic corrections, but these make
little difference for a low spin, especially as the gravitational
redshift more or less cancels the Doppler blueshift at our chosen
inclination of $60^\circ$ for a Schwarszchild black hole (Zhang et
al. 1997).  

The red line in Fig. \ref{fig:optxhub}a shows this full disc spectrum
calculated for a $10^6$\msun Schwarzchild black hole accreting at
$L/L_{Edd}=1$. The peak effective disc temperature of $4\times 10^5$~K
gives $f_{col}=2.34$ from Equation 1 for the innermost radii.  At
larger radii the temperature is lower, so according to Equation 1 the
colour temperature correction should increase. However, this assumes
that electron scattering completely dominates the opacity. This is
certainly not true for $T<10^4$~K, when H is neutral so there are few
free electrons, so $f_{col}=1$ at this point. The blue line in
Fig. \ref{fig:optxhub}a shows a simple colour temperature corrected
blackbody disc with $f_{col}=2.34$ on all annuli with temperature
$T>T_{scatt}=10^5$~K (a conservative point at which electron
scattering should completely dominate the opacity as being well beyond
the ionisation energies of H and He), and $f_{col}=1$ for temperatures
below this. This matches fairly well to the full disc spectrum around
the peak, though the abrupt jump in colour temperature from $2.34$ to
$1$ at $T=10^5$~K gives a sharp change in slope in the EUV spectrum
while the full spectrum calculation (red line) is much smoother,
showing that the colour temperature correction decreases more
gradually with temperature. There is also some residual $f_{col}>1$ in
the optical above the Balmer edge, as shown by the slightly lower
normalisation of the full disc spectrum in this bandpass, though this
effect is much smaller than the potential change in normalisation
arising from different inclination angles (see e.g. Hubeny et
al. 2001).

\begin{figure}
\centering
\leavevmode  \epsfxsize=6.5cm \epsfbox{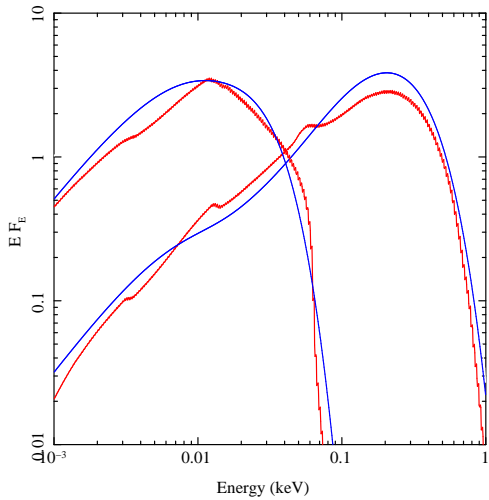}
\caption{Blue lines show Schwarzschild disc spectra calculated using 
colour temperature corrected
blackbodies with $f_{col}$ given by Equations 1 and 2. These compare
well to the full radiative transfer disc spectra (red lines), giving smoother
spectra than those in Fig 1 calculated from a single $f_{col}$.
The lower temperature spectra are for $M=2\times 10^8$\msun
with $L/L_{Edd}=0.05$ while the higher temperature are 
$M=10^6$\msun, $L/L_{Edd}=1$. 
}
\label{fig:optxagnf}
\end{figure}

The value of the colour temperature correction in the general case
where electron scattering is not overwhelmingly dominant is not
possible to estimate analytically, as it depends on the details of the
vertical structure. We estimate this instead from full radiative
transfer disc spectral calculations.  The red lines in
Fig. \ref{fig:optxhub}b show the results of this for Schwarzchild
black holes with mass $10^8$\msun at $L/L_{Edd}=0.2$ (upper lines)
compared to $2\times 10^8$\msun at $L/L_{Edd}=0.05$. The total
luminosity $L=\eta \dot{M}c^2$ is simply set by the total mass
accretion rate $\dot{M}\propto M L/L_{Edd}$ and black hole efficiency,
$\eta$, so is a factor 2 higher for the spectrum with $10^8$\msun at
$L/L_{Edd}=0.2$.  Conversely, the optical luminosity is set by the
combination of $M\dot{M}$ (e.g Davis \& Laor 2011) $\propto M^2
L/L_{Edd}$ i.e. is the same for both models. The full radiative
transfer spectra (red solid lines) again have strong edge features, so
are not well described by a set of (colour temperature corrected)
blackbodies. Nonetheless, these spectra can be roughly modelled with
$T_{scatt}=4\times 10^4$~K and $f_{col}=2$ and $1.8$ for ($M$,
$L/L_{Edd}$) of ($10^8$\msun, $0.2$) and ($2\times 10^8$\msun,
$0.05$), respectively (blue lines). Complete thermalisation
($f_{col}=1$) is shown as the black dashed line. This is not very
different to the full spectrum for $M=2\times 10^8$\msun,
$L/L_{Edd}=0.05$ as the highest temperature in the (blackbody) disc is
only $5.4\times 10^4$~K. Thus He is mainly neutral, reducing the
number density of free electrons, and the absorption opacity is
starting to reach its peak around the ionisation of Hydrogen (see
e.g. Hure et al. 1994). Thus electron scattering is not overwhelmingly
dominant, so Equation 1 overestimates $f_{col}$ at $2.9$.

Davis \& Laor (2011) show that electron scattering is even less
important for the case of a Schwarzchild black hole at $M=10^8$\msun,
$L/L_{Edd}=0.027$ (i.e. $\dot{M}=0.1$\msun yr$^{-1}$), as the
blackbody disc is a fairly good approximation to the full radiative
transfer disc spectrum i.e. the $f_{col}\to 1$ as the disc temperature
drops below $3\times 10^4$~K, where H starts to become neutral so the
associated absorption edge opacity is large.

We can approximate the change in colour temperature correction as
$$f_{col}\sim  (T_{max}/3\times 10^4)^{0.82}\eqno{(2)}$$
over  the critical temperature range of $3\times 10^4~K<T_{max}<10^5~K$.
This changes linearly with $T_{max}$ from unity to 2.7 (the value of
$f_{col}$ from Equation 1 for $T_{max}=10^5$~K).
Electron scattering should dominate above $10^5$~K, so 
$f_{col}$ is then given by Equation 1, and decreases slowly 
with $T_{max}$. 

Thus the
observed maximum disc temperature increases as $T_{max,
obs}=T_{max}f_{col}(T_{max}) \propto T_{max}^{1.82}\propto
[(L/L_{Edd})/M]^{0.46}$
in the range $3\times 10^4~K < T_{max} < 10^5~K$. This is 
much faster than the  $[(L/L_{Edd})/M]^{0.25}$
dependence predicted from purely blackbody
models. Fig. \ref{fig:optxagnf} shows how the disc spectrum can be
fairly well matched to the full radiative transfer models by 
combining Equations 1 and 2 to calculate 
$f_{col}$ at each radius in the disc. 

In summary, the spectrum of a bare accretion disc can be roughly
described by a colour temperature corrected sum of blackbody spectra,
where the value of the colour temperature correction factor is
approximately unity in the optical part of the spectrum, but is higher
above the Lyman break where Hydrogen and Helium are ionised and
electron scattering can dominate. Where scattering is completely
saturated, the colour temperature correction is as large as predicted
by Equation 1, but this only occurs for a disc with intrinsically high
temperature i.e. for high $L/L_{Edd}$ and low black hole mass i.e. for
Narrow Line Seyfert 1s.  This colour temperature correction
accentuates the expected increase in disc temperature for these
objects due to their low mass/high mass accretion rate.  This
increases the distinction in EUV and hence emission line properties
between objects with low $(L/L_{Edd})/M$ (typically Broad Line Seyfert
1 galaxies) and those with high $(L/L_{Edd})/M$ (typical Narrow Line
Seyfert 1s).

\section{Comptonised disc emission for the Soft X-ray Excess}

\begin{figure}
\centering
\leavevmode  \epsfxsize=6.5cm \epsfbox{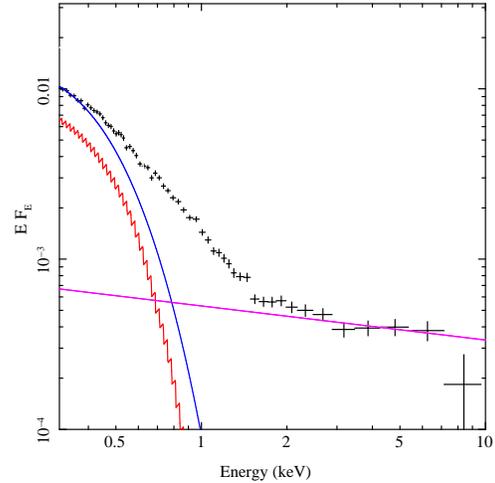}
\caption{The XMM-Newton PN spectrum of REJ1034+396 (black points)
compared to the full radiative transfer disc spectrum for 
$M=10^6$\msun, $L/L_{Edd}=1$ (red), together with the
colour temperature corrected blackbody disc spectrum
for $M=10^6$\msun, $L/L_{Edd}=1.15$. This can account for 
most of the 'soft X-ray excess' in this object, though the 
shape of the
predicted disc spectrum is too steep to match the observed data in the
0.5-1~keV energy range even including 
the best fitting 2-10~keV power law (magenta line).
}
\label{fig:re1034sx}
\end{figure}

Fig. \ref{fig:re1034sx} shows this full radiative transfer AGN spectrum
(red line) compared to the X-ray spectrum of the NLS1 REJ 1034+394
(black data points) assuming a distance of
188~Mpc as appropriate for its redshift of 0.04. This is one of the
most extreme NLS1 systems, with permitted line widths which are
amongst the lowest seen (Grupe et al. 2004). Subtracting the narrow
component of H$\beta$ using the [OIII] line profile gives a residual
which has a FWHM of 988~km/s indicating a black hole mass of $\sim
1.5\times 10^6$\msun when combined with the optical luminosity of
$\lambda L_{5100}=1.9\times 10^{43}$~ ergs s$^{-1}$ (Jin et al. 2011a,
hereafter J11a). The bolometric correction is highly uncertain since
the luminosity peaks in the unobservable EUV (Puchnarewicz et al. 2001;
Casebeer, Leighly \& Baron 2006), but estimates for $L/L_{Edd}$ range from $\sim
1-5$, so the system parameters are directly comparable to the full
radiative transfer disc model. We include the best fitting 2-10~keV
power law spectrum (magenta line), and Galactic absorption column of
$1.47\times 10^{20}$~cm$^{-2}$. Fig. \ref{fig:re1034sx} shows that the
predicted emission from the disc itself, even without Compton
upscattering, will extend into the soft X-ray range for such a low
black hole mass/high mass accretion rate. However, it also illustrates
that the disc spectrum rolls over too rapidly to explain the 0.6-2~keV
emission, so some Compton upscattering is required.

As briefly outlined in the Introduction, the problem is that the soft
excess appears to have a remarkably constant temperature, which seems
finetuned for external Comptonisation models.  Intrinsic disc Compton
scattering gives more scope for a 'thermostat', e.g. enhanced enhanced
dissipation in the upper layers of the disc over and above that
expected from the Shakura-Sunyaev prescription (e.g. Davis et
al. 2005). Enhanced dissipation away from the midplane
is indeed found in self consistent
simulations of the vertical structure produced with the
MRI in a radiation
pressure dominated disc (Blaes et al. 2011).  The effective depth of
the photosphere then controls both $\ell_h/\ell_s$ and $\tau$, such
that seeing deeper into the disc means that there is both additional
dissipation to heat the electrons but also larger optical depth of
electrons. This might result in an energy per electron (i.e.  temperature)
which is fairly constant due to this physical correlation between the
two parameters.

Alternatively (or additionally) the radiation pressure dominated MRI
appears to drive large scale turbulence. This bulk motion, with mean
velocity $<v^2>$, can Compton upscatter the emerging disc flux,
producing the same spectrum as from $3kT\sim 1/3 <v^2>$ (Socrates,
Davis \& Blaes 2004). The apparent finetuning of temperature then
instead requires a constant typical turbulent velocity which should
scale mainly with $\dot{m}$ rather than black hole mass (Socrates,
Davis \& Blaes 2004). The AGN sample of Gierli\'{n}ski \& Done (2004)
mostly had high mass accretion rate objects, so the range in $\dot{m}$
is rather small giving a similarly small range in characteristic
temperature for the bulk Compton upscattering.  We note that this
could also explain why this component does not seem to be present in
BHB, as their much higher temperature discs means that it would be
unaffected by the typical $kT\sim 0.2$~keV of the soft excess
component seen in AGN. 

\begin{figure}
\centering
\leavevmode  \epsfxsize=6.5cm \epsfbox{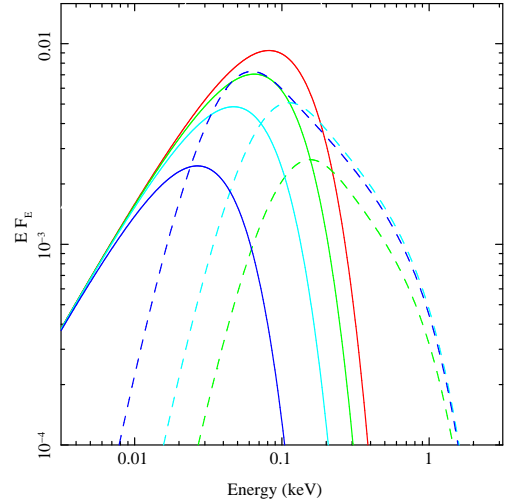}
\caption{The model spectra for a disc where the accretion flow 
thermalises into blackbody emission with $f_{col}=1$ 
for radii $r > r_{corona}$ (solid line). Below this radius the
energy is dissipated 
by Compton upscattering of seed photons at the disc temperature
at $r_{corona}$ by electron with temperature of 0.22~keV and optical
depth of 15 (dotted line). The red, green, cyan and blue models
are calculated for $r_{corona}=6, 16, 30$ and $75$, corresponding
to 0 (pure disc), 25, 50 and 75\% of the total accretion 
power in the Compton upscattering region. 
}
\label{fig:rcorona}
\end{figure}

Further possibilities include changes in disc structure due to
advection of radiation at high mass accretion rates (slim discs,
Abramowicz et al 1989). The disc becomes so optically thick that
photons are swept radially along with the flow before they can diffuse
outwards. However, the rapid increase in velocity in the plunging
region means that the flow can become optically thin 
at this point, so some of the radiation can escape from within the
innermost stable circular orbit (Sadowski 2009) 

Whatever the origin, a simple approach to fitting the soft X-ray data
is to assume that there is an additional Compton upscattered component
e.g.  the {\sc xspec} model {\sc compTT} (Titarchuk 1994) as well as
the optical/UV emitting disc.  However, it is difficult to uniquely
determine the parameters of two independent components especially
given the lack of data in the EUV region (see e.g. Jin et
al. 2009). More fundamentally, the Comptonised emission should also be
powered ultimately by the accretion flow, so the two components should
be energetically coupled.

Done \& Kubota (2006, hereafter DK06) developed a model including this
energetic coupling for the strongly Comptonised very high state of
Galactic Black Hole Binaries (BHB). Here we develop a version of this
model which is much faster (so is more appropriate for fitting
multiple spectra), and which explicitally uses the physical parameters
as inputs (see the Appendix for a comparison of the two models).

We assume that this Compton upscattering takes place in the disc
itself, so the luminosity in this Comptonisation component should be
the same as that of the disc luminosity
at each radius. Thus once the mass and mass accretion rate through the
outer disc are set, the total luminosity of the Compton upscattered
flux is completely determined by the integral of the emissivity from
$r_{corona}$, the radius at which the emission starts to emerge as
Comptonised rather than (colour temperature corrected) blackbody, to
$r_{isco}$, the last stable orbit. We assume that the energy within
$r_{corona}$ is emitted as Compton upscattered flux ({\sc comptt}),
with seed photons characterised by the (colour temperature corrected) 
disc temperature at $r_{corona}$.

We illustrate the dependence of the spectrum on
$r_{corona}$ in Fig. \ref{fig:rcorona} for our fiducial system (black
hole mass of $10^6$~\msun at $L/L_{Edd}=1$ calculated with
$f_{col}=1$). We show results for $r_{corona}=$~6 (i.e. same as the
black dashed line in Fig. \ref{fig:optxhub}a) 16, 30 and 75,
corresponding to 0, 25, 50 and 75\% of the total accretion power
dissipated in the corona, respectively (red, green, cyan and blue
lines).  The solid lines show the emission from the outer disc, while
the dotted lines show the increasing power in the soft Comptonisation
component, all of which have $kT_e=0.22$~keV and $\tau=15$.  For these
parameters the Comptonised spectrum is steep, so it peaks at the seed
photon temperature which decreases with increasing $r_{corona}$, as is
also seen in the lower maximum temperature emission from the blackbody
disc emission.  The optical/near UV region is dominated by the outer
disc emission, so is the same for all the models considered here.

As discussed in the Appendix, our assumption of the seed photon energy
as being set by the disc temperature at $r_{corona}$ may not be
accurate for very large $r_{corona}$.  Steep Comptonised spectra peak
at ($\sim 4\times$) their seed photon temperature (see
Fig. \ref{fig:rcorona}), so the spectrum can peak at energies lower
than that of the thermalised disc for $r_{corona}>75$ which is
probably unphysical. Seed photons generated internally in the
Comptonisation region probably are more important, and these would
have higher temperature. We stress that the derived parameters should
be used with caution where $r_{corona}$ is large and the
Comptonisation is steep.

\section{Modelling the SED of a NLS1: REJ0134+396}

\begin{figure}
\centering
\leavevmode \epsfxsize=6.5cm \epsfclipon \epsfbox[400 400 700 600] {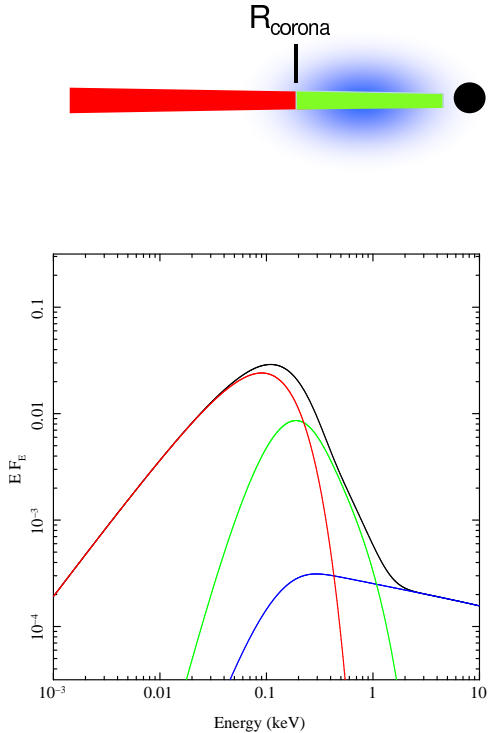}
\leavevmode \epsfxsize=6.5cm \epsfbox{optxagn_comp3.ps}
\caption{A schematic of the model geometry and resultant spectra, with
outer disc (red) which emitts as a (colour tempearture corrected)
blackbody, and an inner disc (green) where the emission is instead
Compton upscattered (perhaps by bulk turbulent motion in the disc,
or by there being more dissipation in the effective photosphere than
assumed in the standard Shakura-Sunyaev vertical dissipation
profile). Some fraction of the energy is also Compton upscattered in
a corona (blue) to produce the power law tail to high energies.
}
\label{fig:model}
\end{figure}

While this gives a model for the soft excess, there is also higher
energy emission above 2~keV (see Fig. \ref{fig:re1034sx}). The energy
to power this must also ultimately be derived from mass accretion, so
we assume that some fraction, $f_{pl}$, of the energy dissipated from
$r_{corona}-r_{isco}$ powers this high energy
Comptonisation component. We model this using {\sc nthcomp} code in
{\sc xspec} (Zdziarski, Johnson \& Magdziarz 1996), 
again setting the seed photon temperature equal to that
of the (colour temperature corrected) blackbody from the disc at
$r_{corona}$. 

Thus the full model contains three distinct spectral components, but
these are all powered by the energy released by a single accretion
flow of constant mass accretion rate, $\dot{M}$, onto the black hole
of mass M. The outer disc emission is the colour temperature corrected
blackbody emission from $r_{out}$ to $r_{corona}$, where $f_{col}$ is
only applied on annulii with temperature $T>T_{scatt}$.  A fraction
$f_{pl}$ of the remaining energy emitted as the mass accretes from
$r_{corona}$ to $r_{isco}$ is emitted as the high energy (above 2~keV)
Comptonisation, characterised by a power law of photon index $\Gamma$
and electron temperature fixed at $100$~keV, while the remaining
fraction ($1-f_{pl}$) is emitted as low temperature, optically thick
Comptonisation of the disc emission, parameterised by $kT_e$ and
$\tau$. Fig. \ref{fig:model} shows a schematic of these three regions,
together with their spectra. We call this full model {\sc optxagn},
and have made it publically available as a local model for the {\sc
xspec} spectral fitting package (Arnaud et al. 1996).  We also include
an additional model, {\sc optxagnf}, incorporating the approximate
$f_{col}(T_{max})$ relationship of Equations 1 and 2. In both
versions, the overall normalisation is set explicitally by the model
parameters, so the normalisation must be frozen to unity. Inclination
angle could change this by a factor $\sim 2$ (see e.g. Hubeny et al
2001), but this is beyond the scope of the model.

We fit the full model to the broad band spectral energy distribution
(SED) assembled by J11a, including absorption by two gas columns, one
fixed at the Galactic value along the line of sight and the other
allowed to be free to model intrinsic absorption associated with the
host galaxy and/or nucleus. Each of these two gas columns is assumed
to have a standard gas to dust ratio, $E(B-V)=1.7 (N_H/10^{22} {\rm
cm}^{-2})$, producing optical reddening which is tied to the X-ray
absorption (see J11a).

\begin{figure}
\centering  
\leavevmode  \epsfxsize=6.5cm \epsfbox{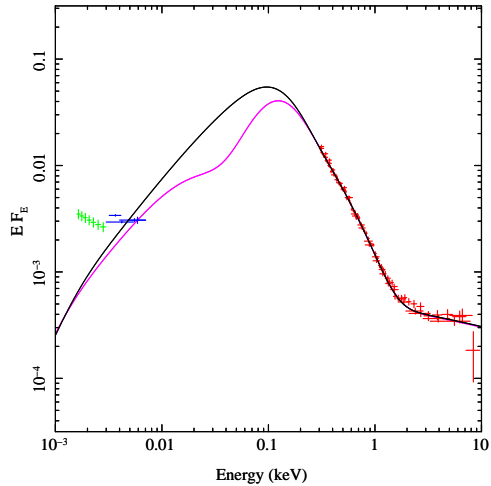}
\caption{The XMM-Newton X-ray (red) and Optical Monitor (blue) data
from RE J1034+396, together with (non-simultaneous) SDSS optical
continuum points (green) from J11a. These are shown (absorption corrected) 
with the best fit models with $f_{col}=1.0$ (black) and 2.6
(magenta). The fit only includes the highest frequency UV (UVW2) point as
HST data show that the disc only dominates in the far UV. The models
differ mainly in the unobservable EUV, but $L/L_{Edd}$ differs only by
a factor of 2 and all other model parameters are similar (see Table 1).}
\label{fig:re1034}
\end{figure}

We allow a range of black hole mass, corresponding to the FWHM of the
intermediate and broad components of H$\beta$ (918 and 4400 km/s
giving $1.2\times 10^6< M < 2.9\times 10^7$\msun, respectively)
Fig. \ref{fig:re1034} shows the best fit, absorption corrected, model
together with the data for REJ1034+396 assuming $f_{col}=1$ (black, as
assumed in J11a) and for $f_{col}= 2.4$, $T_{scatt}=10^5$~K (magenta, as in
Fig 1).
We show the full optical/UV SED points from J11a but 
only include the highest frequency UV point in the fit (XMM-Newton 
OM UVW2) as the
continuum emission has a very different spectrum to that expected from
a disc even in the UV (Casebeer, Leighly \& Baron 2006).  
This is probably due to there being an 
additional continuum component, most probably starlight contamination from
the host galaxy (Sani et al. 2010).

Fig. \ref{fig:re1034} shows that the models differ in the unobservable
EUV regime, but both are consistent with the UV flux limits
for the disc emission, and both fit the soft X-ray data well, with
best fit parameters given in Table \ref{tab:fit1034}. The spectral
slope of the soft X-ray excess sets the Comptonisation parameters,
$kT_e$ and $\tau$, and extends down to the seed photon energy, set
by the inner disc temperature at $r_{corona}$. This is where the
luminosity peaks for a steep soft X-ray excess, and so the
substantially higher seed photon temperature arising from
$f_{col}=2.6$ compared to $f_{col}=1$ results in a substantially lower
total luminosity. The accretion efficiency, $\eta$, is fixed by black
hole spin (assumed to be 0), so the total luminosity $L=\eta \dot{M}c^2$ sets the total
mass accretion rate. Conversely, the optical luminosity is set by the
combination of $M\dot{M}$ (e.g Davis \& Laor 2011), so for a higher
$\dot{M}$ to fit the same optical/UV flux points requires a lower
black hole mass. Thus the model with $f_{col}=1$ has both a higher
$\dot{M}$ and lower $M$ (and hence much higher $L/L_{Edd}$) than the
corresponding model with $f_{col}=2.4$, but none of the other
parameters change significantly.

\begin{table*}
\begin{tabular}{lc|l|l|l|l|l|l|l|l|l|l}
 \hline
   \small $N_H$ ($10^{20}$~cm$^{-2}$) & $f_{col}$ & $M$ ($10^6$\msun)
   & $L/L_{Edd}$ & $r_{corona}$ ($R_g$)& $kT_e$ (keV)& $\tau$ &  
 $\Gamma$ & $f_{pl}$ & $\chi^2/\nu$ \\ 
\hline
$1.7\pm 0.7$ & 1.0 & $1.2^{+0.1}$ & $5.0_{-0.6}^{+0.7}$ & $31_{-9}^{+14}$ &
$0.23\pm 0.03$ & $11\pm 1$ & 2.2 & $0.05\pm 0.02$ & 631/563\\
$1.6\pm 0.6$ & 2.4 & $1.9_{-0.1}^{+0.8}$ &  $2.4_{-0.6}^{+0.3}$ & $100^{+*}_{-60}$ & 
$0.23\pm 0.03$ & $11\pm 1$ & 2.2 & $0.05\pm 0.02$ & 630/563\\
\hline
\end{tabular}
\caption{Details of fits to the NLS1 RE1034+396 using the 
intrinsic Comptonisation model for the soft excess. The first line has
$f_{col}=1.0$ while the second is for $f_{col}=2.4$. 
}
\label{tab:fit1034}
\end{table*}

\section{Modelling the SED of a BLS1: PG1048+342}

\begin{figure}
\centering  
\leavevmode  \epsfxsize=6.5cm \epsfbox{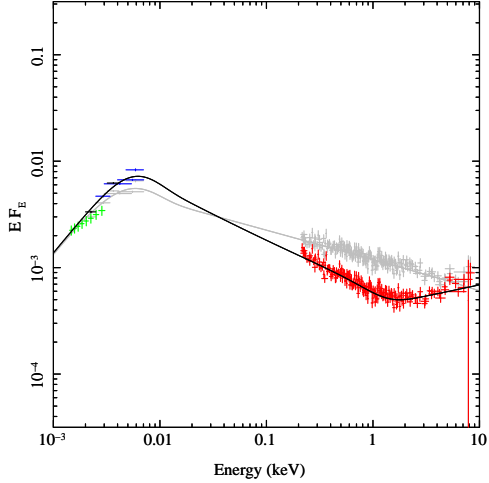}
\caption{As for Fig. \ref{fig:re1034} but for the broad line Seyfert 1
PG1048+213. The best fit model with $f_{col}=1.0$ (black) is shown,
but this is identical to one with $f_{col}=2$ as the maximum disc
temperature at $r>r_{corona}$ is always less than $T_{scatt}=4\times
10^4$~K. The grey line and data show the different resulting SED
derived from assuming that there is complex absorption producing the
dip in the X-ray spectrum.  Nontheless, both models give a similarly
large non-disc component.  
}
\label{fig:pg1048}
\end{figure}

While the disc almost certainly contributes to the soft excess in low
mass, high mass accretion rate AGN such as RE J1034+396, the same
should not be true for the higher mass, lower mass accretion rate
broad line AGN as the disc temperature is lower. This temperature is
too low for electron scattering to ever dominate so there is no
substantial colour temperature correction (see
Fig. \ref{fig:optxhub}b), pulling the intrinsic disc emission even
further away from the soft X-ray region. Nonetheless, these objects
still show a soft X-ray excess. We illustrate this using the SED of
PG1048+213, which has a black hole mass of $\sim 2\times 10^8$\msun,
and $L/L_{Edd}\sim 0.1$ (J11a).  Fig. \ref{fig:pg1048}a shows the
deconvolved spectrum for $f_{col}=1.0$, and it is clear that not only
is the disc intrinsically cool, but that the fit requires a very large
$r_{corona}$ in order to produce the observed X-ray emission (both
soft excess and higher energy power law), so that only the coolest
outer disc contributes to the themalised emission.

\begin{table*}
\begin{tabular}{lc|l|l|l|l|l|l|l|l|l}
 \hline
   \small $N_H$ ($10^{20}$~cm$^{-1}$) & $f_{col}$ &
$L/L_{Edd}$ & $r_{corona}$ ($R_g$)&
$kT_e$ (keV)& $\tau$ &   $\Gamma$ & $f_{pl}$ & $\chi^2/\nu$ \\ 
   \small & &  & & $N_H$ & $C_f$  \\ 
\hline
$1.9\pm 0.5$ & 1.0 & $0.17\pm 0.01$ & $100_{-4}$ & 
$0.25\pm 0.05$ & $15\pm 1.5$ &  $1.80\pm 0.08$ &  $0.25\pm 0.02$ & 832/596 \\
$0^{+1}$ & 1.0 & $0.15\pm 0.01$ & $77\pm 9$ & $17\pm 6$ & 
$0.51\pm 0.03$ & $2.27\pm 0.02$ & 1.0 & 884/597\\
\hline
\end{tabular}
\caption{Details of fits to the BLS1 PG1048+213. The first fit is for
the intrinsic Comptonisation model for the soft excess, the second is
for an atomic absorption (partial covering) origin for this
feature. While these are both calculated for $f_{col}=1.0$, models
with $f_{col}=2$ give identical fits as the disc temperature
for $r>r_{corona}$ never rises above $T_{scatt}=4\times 10^4$~K.
}
\label{tab:pg1048}
\end{table*}

While the model fits the data well, it illustrates some of the more
puzzling issues associated with the soft X-ray excess. While turbulent
motions may well give rise to bulk Comptonisation, the mass accretion
rate here is approximately $10\times$ lower in terms of Eddington
fraction than for REJ1034+396, yet the electron temperatures for the
soft Comptonisation are very similar.  Other issues are that the
2-10~keV power law is quite flat with $\Gamma=1.7$ i.e. rising in $\nu
f_\nu$. This looks much more similar to the low/hard state in black
hole binary systems, yet these are seen only at much lower Eddington
fractions ($L/L_{Edd}\la 0.02$) except in the non-equilibrium states
triggered by the dramatic outbursts in Galactic binary systems
(Nowak 1995; Maccarone 2003; Gladstone, Done \& Gierli\'{n}ski 2007; 
Yu \& Yan  2009). The clear disc
component in the optical/UV SED and its dominance over the level of
X-ray emission makes this appear more similar to the high/soft state
in Galactic black hole binaries, but these have $\Gamma=2-2.2$ (see
e.g. the review by Done, Gierli\'{n}ski \& Kubota 2007) (hereafter DGK07).

These issues have motivated alternative ways to model the soft X-ray
excess, as a distortion from atomic processes. There is a strong jump
in opacity at ~0.7 keV from partially ionized material, where
OVII/OVIII and the Fe M shell unresolved transition array (UTA)
combine to produce much more absorption above this energy than
below. This could produce an {\em apparent} (rather than intrinsic)
soft excess in two very different geometries, either by reflection
from optically thick material out of the line of sight (Fabian et al
2002), or through absorption by optically thin material in the line of
sight (Gierli\'{n}ski \& Done 2004; Chevallier et al. 2006). However, both
these geometries would show characteristic sharp atomic features
(lines and edges).  These can be smeared by strong relativistic
effects in the reflection model, but the parameters required are quite
extreme (the disc has to extend down to the last stable orbit of a
high spin black hole, with reflection emissivity which is highly
concentrated to the innermost regions: Crummy et al. 2006). In
absorption, the line of sight velocity shear required to smear out the
atomic features is probably unrealistically large (Gierli\'{n}ski \& Done
2004; Schurch \& Done 2007).  However, the absorption model may be
substantially more complex as the high columns inferred mean that
resonance line scattering should be important (Sim et al. 2010),
and the absorption can be clumpy so that it covers only part of the
source, diluting the expected line absorption signature (Miller et al
2007; 2009; Turner et al. 2007; Risaliti et al. 2011).

Given the limited signal to noise of our X-ray data, we model the
effects of absorption and/or reflection using a simple neutral partial
covering model.  This assumes that some fraction, $C_f$ of the X-ray
source is covered by a neutral column of $N_H$, while the rest is
unobscured. Since this model can match the whole X--ray spectrum with
a single power law, there is no need for a separate soft excess
component. Hence we assume that all the gravitational energy released
from inside $r_{corona}$ goes into powering the high energy tail. 
The grey data and model line in Fig. \ref{fig:pg1048} shows
the resulting inferred SED. While this is somewhat different, the
fraction of luminosity in the EUV/X-ray region is similarly high, so
$r_{corona}$ is still required to be 
large in order to power this emission.  The
inferred luminosity of the Comptonised spectrum after accounting for
distortion by complex absorption still contains $3\times$ more power
than that which thermalises. If this is powered by material that
accretes through the outer thin disc then this requires all the
gravitational energy from within $\sim 75~R_g$. Unless the hard X-rays
are powered by a separate coronal flow, the inescapable conclusion is
that the accretion energy cannot thermalise in a thin disc structure
down to the last stable orbit. Such a geometry is strongly reminiscent
of the truncated disc models for the low/hard state in stellar mass
black hole binary systems, yet the Eddington fraction of $\sim 0.15$
is relatively high for this state (see e.g. the review by DGK07).

Both models in Table 2 have large $r_{corona}$ and a steep 
Comptonisation spectrum. This means that the full SED peaks at 
($\sim 4\times$ the) seed photon temperature for Comptonisation,
set by the disc temperature at $r_{corona}$. This probably
underestimates the  seed photon temperature, as discussed in 
Section 3, so the (unobservable) EUV part of the spectrum could
have somewhat different shape (see Appendix).

\section{Modelling the mean optical to X-ray SED of AGN}

\begin{table*}
\begin{tabular}{lc|l|l|l|l|l|l|l|l}
 \hline
   \small spectrum & $L/L_{Edd}$ & $M$\msun & $f_{col}$ & $r_{corona}$ ($R_g$) &
$kT_e$ (keV) & $\tau$ &   $\Gamma$ & $f_{pl}$  \\ 

\hline
M1 & 0.049 & $1.38\times 10^8$ & 1 & $60\pm 5$ &
$0.17_{-0.02}^{+0.03}$  & $21\pm 2$ & $1.80\pm 0.02$ & $0.52\pm 0.02$ \\
M2 & 0.23 & $1.15\times 10^8$ & 1 & $42_{-6}^{+15}$ &
$0.31_{-0.08}^{+0.15}$  & $13\pm 3$ & $1.87\pm 0.02$ & $0.34\pm 0.04$ \\ 
M3 & 2.1 &  $1.55\times 10^7$ & 1 & $20\pm 3$ & $0.54_{-0.23}^{+0.60}$
& $6.7_{-5.5}^{+2.3}$ & $2.03\pm 0.02$ & $0.11\pm 0.02$ \\ 
\hline
M3 & $0.77\pm 0.06$ & $2.6_{-0.2}^{+0.4}\times 10^7$ & 2.5 & $12_{-2}^{+8}$ & 
$0.26_{-0.10}^{+0.29}$ & $12_{-6}^{+11} $ & $2.03\pm 0.02$ &
$0.35_{0.06}^{+0.15}$ \\
\hline
\end{tabular}
\caption{Details of fits to the three mean spectra in J11b.
}
\label{tab:mean}
\end{table*}

\begin{figure*}
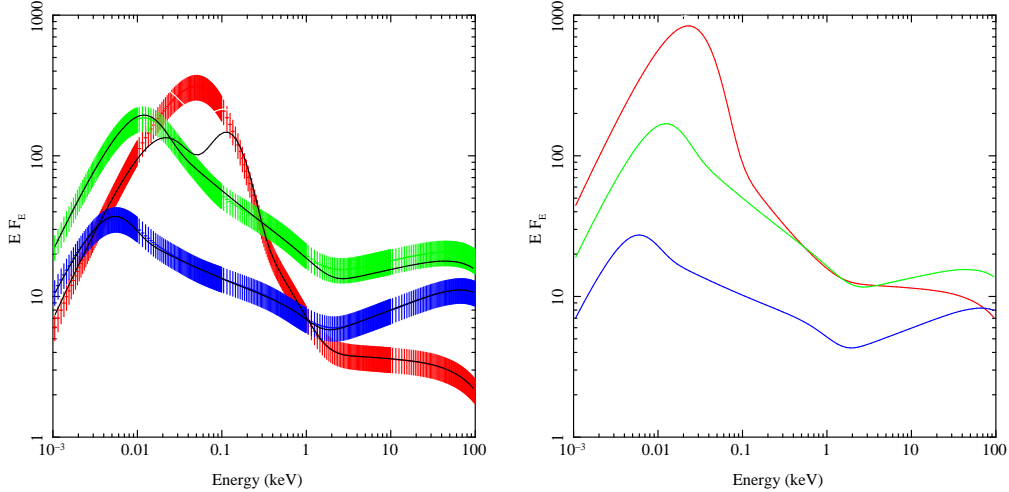

\centering  
\leavevmode  
\begin{tabular}{cc}
\epsfxsize=6.5cm \epsfbox{mean_fcol25_new_3.ps} &
\epsfxsize=6.5cm \epsfbox{mean_mass_1e8_3.ps}\\
\end{tabular}
\caption{a) The three mean spectra from J11b, derived using
$f_{col}=1$, but we show the fit (excluding the unobervable
0.01-0.3~keV region) with $f_{col}=2.5$ for the lowest mass/highest
mass accretion rate spectrum, where the disc temperature 
exceeds $T_{scatt}=10^5$~K. M1 (blue) has $L/L_{Edd}=0.058$
and  black hole mass of $1.4\times 10^8$\msun. M2 (green) has 
$L/L_{Edd}=0.23$ and black hole mass of $ 1.1\times 10^8$\msun. 
M3 (red) has $L/L_{Edd}=0.77$ 
and black hole mass of $2.6\times 10^7$\msun. b) shows the spectral
evolution with $L/L_{Edd}$ alone by redoing each model for 
a single black hole mass of $10^8$\msun (and $f_{col}=1$).
}
\label{fig:mean}
\end{figure*}

Having discussed the different issues in fitting a single NLS1 and
BLS1 spectrum, we now use the mean spectra as compiled by J11a and Jin
et al. (2011b), (hereafter J11b) from stacking the derived spectral
models from fitting their sample of 51 objects into 3 subgroups.  Of
the various possible parameters to group by (e.g. FWHM of H$\beta$,
black hole mass etc) $L/L_{Edd}$ gave the minimum dispersion, so is
the most likely to represent the principle underlying physical
parameter which determines the spectral shape (J11b).  We refer to
these spectra as M1, M2 and M3 for $L/L_{Edd}$ of 0.05, 0.2 and 2.1,
respectively.

These three spectra have an associated mean black hole mass which is
anti-correlated with $L/L_{Edd}$.  It is likely that this is primarily
a selection effect.  The optical luminosity of the accretion flow is
$\propto (M\dot{M})^{2/3}$ (e.g. Davis \& Laor 2011) $\propto [M^2
(L/L_{Edd})]^{2/3}$.  This has to be large enough to be detected above
the bulge luminosity of the host galaxy $L_{host}\propto
M_{host}\propto M$ (Magorrian et al 1998). This sets a lower limit on
$M(L/L_{Edd})^2$ for the AGN to be detectable. Thus low mass black
holes can only be detected against their (low mass) host galaxy at
high Eddington fractions, while high mass black holes (hosted by more
massive galaxies) can be detected at much lower Eddington
fractions. These high mass black holes would be even more obvious at
high Eddington fractions, but such objects are very rare in the local
Universe firstly as high mass black holes are much less numerous than
low mass ones, and secondly as their large host galaxies form in the
most overdense regions, where activity (both star formation and
accretion flows onto the nucleus) peaks at redshift $\sim 2$ then
declines as the gas supply runs out. Thus the highest mass black holes
in the local Universe typically have low mass accretion rates
(downsizing, see e.g. Fanidakis et al. 2010).

We show these mean spectra in Fig. \ref{fig:mean}, where the overall
normalisation is arbitrary but is fixed to the same value for all 3
spectra.  The lowest $L/L_{Edd}$ spectrum (M1: blue) has disc to hard X-ray
ratio which is typical of Seyferts in the local Universe (Koratkar \&
Blaes 1999). The middle spectrum (M2: green) is similar but has a slightly
stronger disc component, matching to the mean radio-quiet Quasar
template of Elvis et al. (1994), but the highest $L/L_{Edd}$ template
(M3: red) is very different in terms of shape. It has the lowest 2-10~keV
X-ray flux, so these objects are systematically missed in samples
based on this energy band. M2 has the highest flux in all observable
wavebands, so these are the objects which will be preferentially
picked up in optical and X-ray surveys. 

We fit the model to the spectral templates to derive the mean
parameters. The spectral templates are made by weighting the
individual spectra (derived with $f_{col}=1$) by luminosity distance,
whereas the mean mass and $L/L_{Edd}$ quoted in J11b is
unweighted. Hence the best fit system parameters for the template
spectra can differ slightly from those of J11b. Since black hole mass
and mass accretion rate are degenerate, we fix the black hole mass to
the unweighted mean values of J11b. The dispersion in mass accretion
rate is smallest for M2 (J11b), so we fix $L/L_{Edd}=0.23$ for M2, and
normalise all the other spectra to this to derive the best fit mean
$L/L_{Edd}$ for M1 and M3. These are only slightly different to the
mean values given in J11b. Table \ref{tab:mean} gives all the
resulting parameters.

We then assess the impact of a colour temperature correction.  
M3 has a maximum disc temperature of $2.5\times 10^5$~K, so we use
Equation 1 to predict $f_{col}=2.5$ in the scattering dominated regime i.e.
above $T_{scatt}=10^5$~K. M1 and M2 are similar in mass and mass
accretion rate to the spectra shown in 
Fig. \ref{fig:optxhub}b, so should have 
$f_{col}\sim 1.8$ and $2.0$ respectively above $T_{scatt}=4\times 
10^4$~K. The majority of the disc emission has $T<T_{scatt}$,
so this only affects the inner disc. 
However, in our model the
energy from the inner disc is required to power the X-ray emission, so
will not be emitted as a blackbody. The considerable X-ray luminosity
requires $r_{corona}$ to be substantially larger than the innermost
stable orbit, which reduces the maximum disc temperature. 
Thus the colour temperature correction has little or
no effect on these spectra, so we only explore its effect on M3.

We refit M3 with $f_{col}=2.5$ for $T>10^5$~K, using the same
distance as before, ignoring the unobservable 0.01-0.3~keV
region. Guided by the results of Section 4 we allow both mass and mass
accretion rate to be free parameters. Again we find that the black
hole mass increases and mass accretion rate decreases, giving a mean
$L/L_{Edd}=0.8$ rather than $2.1$ as found for $f_{col}=1$, but all
other parameters are consistent with their previous values (see last
line of Table \ref{tab:mean}).

Since the model is based on physical parameters, we can reset the
black hole mass to the same value to explore the impact of changing
$L/L_{Edd}$ alone. We choose a fiducial mass of $10^8$\msun (very
similar to that of M1 and M2) and set the mass accretion rate to give
the same $L/L_{Edd}$ as that derived for each of the spectra in Table
\ref{tab:mean} (using the parameters from the colour temperature
corrected fits to M3).  The solid lines in Fig. \ref{fig:mean}b show
the resulting evolution of spectral shape in AGN with $L/L_{Edd}$
alone.

The spectral parameters show a clear evolution with increasing
$L/L_{Edd}$, most obviously with $r_{corona}$ decreasing, while the
2-10~keV spectral index increases. Changes in the other parameters are
not so large, but there is a systematic increase in temperature of the
soft excess component, together with a correlated decrease in optical depth
so that its spectrum steepens. The amount of power in the soft
excess compared to the higher energy emission drops slightly, but the
decrease in $r_{corona}$ means that both hard and soft 
X-ray components carry a smaller
fraction of the bolometric luminosity (see also Vasudevan \& Fabian
2007). 

\section{Analogy with Black Hole Binary systems}

BHB show (at least) three different states, variously termed the
low/hard, high/soft (a.k.a thermal dominant or disc dominant) and very
high (a.k.a. steep power law state). Spectra from this latter state
appear very similar to intermediate spectra seen as the source makes a
transition from the low/hard to high/soft states (e.g. Belloni et al
2005).  It is plain from Fig. \ref{fig:mean}b that the highest
$L/L_{Edd}$ spectrum of M3 is very much like a disc dominated
high/soft state spectrum with a small amount of additional soft
Comptonisation. The lower luminosity spectra are more ambiguous,
and could be either low/hard state spectra or intrinsically steeper 
very high state spectra
which are modified by complex absorption/reflection. We explore each
of these possiblities below, and then discuss how we can distinguish
between the two scenarios.

\subsection{AGN sequence as a low/hard to high/soft transition}

The increasing dominance of the disc with increasing $L/L_{Edd}$,
together with the switch from hard to soft 2-10~keV X-ray emission is
initially very reminiscent of the spectral transition seen in stellar
mass galactic black hole binary systems (BHB).  These show a clear
switch from a hard power law dominated low/hard state to disc
dominated high/soft state with a steep tail to higher energies (see
e.g. the reviews by McClintock \& Remillard 2006 hereafter MR06; DGK07).
This is generally
interpreted as a decreasing transition radius between a cool disc and
hot comptonised region (Esin, McClintock \& Narayan 1997), 
matching the decreasing
$r_{corona}$ seen in our AGN fits.

The BHB in the low/hard state also often show a complex continuum,
with a softer Comptonisation component from the outer parts of the
flow which intercept more seed photons from the disc, together with a
harder tail from the hotter central regions (Ibragimov et al 2005;
Takahashi et al. 2008; Makishima et al. 2008; Kawabata \& Mineshige
2010). This inhomogeneous Comptonisation in BHB is also required to
explain the time lags seen in the data (Kotov, Churazov \& Gilfanov
2001; Ar\'{e}valo \& Uttley 2006).  Thus the two component (soft and
hard) Comptonisation required to fit the X-ray spectra in M1 and M2
could be analogous to the two component Comptonisation spectra seen in
the BHB low/hard state.

However, there are also some significant differences.  The transition
in BHB occurs at $L/L_{Edd}\sim 0.02$ when the mass accretion rate is
slowly declining, (Maccarone 2003). This is as expected for an
advection dominated flow, which collapses when the energy transfer
between the ions and electrons becomes efficient. The collapse 
depends only on
optical depth of the flow rather than black hole mass, so should be
the same for both AGN and BHB (Narayan \& Yi 1995).  Yet in these AGN
the low/hard state must persist up to at least $L/L_{Edd}\sim 0.2$ to
explain M2.  Such high transition luminosities are only seen in BHB
during the rapid rise during disc outbursts (see e.g. the compilation
by Yu \& Yan 2009) which drive the flow out of its equilibrium states
(Gladstone, Done \& Gierli\'{n}ski 2007). Our M2 spectrum would then be
analogous to the intermediate state seen during this transition, but
this transition is very rapid in BHB so these intermediate spectra are
rare (Dunn et al. 2009). By contrast, M2 is very similar to the mean
QSO spectral template of Elvis et al. (1994). and has similar
mass and mass accretion rate as a typical QSO (Kollmeier et al. 2006;
Steinhardt \& Elvis 2010) so it must be a very common state. Thus
it does not seem very likely that all these AGN are preferentially
seen during a dramatic rise in mass accretion rate. The only real
possibility for this AGN sequence to represent a low/hard to high/soft
transition is if there is some weak mass dependence on the critical
luminosity  not predicted by the advective flow models (Narayan \& Yi
1995).

\subsection{AGN sequence as high mass accretion rate transition}

This motivates us to explore the alternative possiblity, that the AGN
spectra seen here are above the low/hard state transition luminosity,
so correspond to one of the high mass accretion states in BHB. The
hard X-ray tail ($\Gamma<2$) is incompatible with this, as these high
mass accretion states in BHB almost always have $\Gamma>2$ (MR06;
DGK06). Hence the observed hard X-ray spectra in M1 and M2 would have
to be distorted by complex absorption and/or reflection, whereas M3
already has a soft 2-10~keV spectrum, so does not require complex
absorption.  This potential difference is supported by the behaviour
of the soft X-ray excess. In the highest $L/L_{Edd}$ spectra (M3 and
REJ1034+396) the soft excess appears as a true excess over a
$\Gamma=2-2.2$ extrapolation of the 10~keV flux level down below
2~keV, unlike M2, M1 and PG1048+231 where the $\nu f_\nu$ flux level
at 0.1~keV is roughly similar to that at 10~keV.  The soft excess seen
in the highest $L/L_{Edd}$ spectra could then represent a 'true' soft
X-ray excess connected to the disc as its approaches Eddington
(perhaps bulk motion Comptonisation from turbulence: Socrates et al 2004, 
or trapped radiation 
advected along with the flow which can then be released in 
the plunging region: Sadowski 2009), while the 'bend'
seen in the lower $L/L_{Edd}$ is a 'fake' soft excess, where the
apparently hard 2-10~keV spectrum and steeper 0.3-2~keV excess are
both distortions from complex absorption and/or reflection.

Complex absorption seems quite likely in an AGN environment, as a UV
bright disc is very efficient in producing a strong wind from UV line
driving (Proga, Stone \& Kallman 2000). The wind should become
stronger as $L/L_{Edd}$ increases, which is at first sight
inconsistent with the requirement that the spectral distortion is
larger in M1 and M2 than in M3. However, the wind also depends
strongly on black hole mass as it is launched from the region where
the disc temperature is around the energies of the strong UV resonance
lines, as this is where the opacity peaks. The increase in UV opacity
for a lower temperature (i.e.  higher mass, lower $L/L_{Edd}$) disc
may more than compensate for the lower $L/L_{Edd}$. The wind mass loss
rate may even be substantial enough to modify the disk structure,
significantly reducing the mass accretion rate below the wind
launching point. We caution that this may require new disc models,
which allow the mass accretion rate to change with radius.

The fraction of luminosity emitted in the inferred $\Gamma\sim 2.2$
tail is substantial in both M1 and M2. These spectra would then
correspond to the very high/steep power law state seen in BHB
(e.g. DK06), whereas M3 would still be a disc dominated state with a
small additional Comptonisation component.

However, the M1/M2 spectra do not appear to be modified by complex
X-ray absorption. While the signal-to-noise in PG 1048+213 is not
overwhelming, this spectrum (Fig 7) is very similar to the much better
data from Mkn 509 which has similar mass aand mass accretion rate
(Mehdipour et al 2011; Noda et al 2011). Here the 'soft excess'
clearly has different variability to the 'power law', supporting a
true two component interpretation of the spectrum (Noda et al
2011). It is the higher mass accretion rate, low mass objects (more
like M3) for which the complex spectral variability is most often
interpreted in terms of reflection and/or complex absorption (Fabian
et al 2002; Fabian et al 2009; Ponti et al (2010); 
Miller et al 2007; 2009; 2010; 2011; Turner et al 
2007).

\subsection{Distinguishing between a low/hard and very high state}

The shape of the power spectra of the rapid X-ray variability
correlates with spectral state in BHB. However, both low/hard and very
high states have variability power spectra which can be roughly
characterised as band limited noise, i.e.  have power spectra with
both a low and high frequency break (see e.g. MR06; DGK07).
Conversely, stationary high/soft states (in Cyg X-1) show only a high
frequency break (e.g. MR06; DGK07). Thus both
possibilities predict that the objects contributing to M2 should
predominanly have power spectra which are band limited noise, while
those which contribute to M3 should extend unbroken to low
frequencies.  Currently there are no objects in our samples with well
defined variability power spectra, but objects with similar
$L/L_{Edd}$ to those in M2/M3 typically show only a high frequency
break i.e. are more similar to the high/soft state in Cyg X-1 (see
e.g. the review by McHardy et al. 2010). However, we caution that if
M1 and M2 do indeed correspond to very high state spectra, distorted
by complex absorption/reflection, then their variability will also be
similarly distorted. Variable obscuration in a clumpy absorber will
add to the intrinsic variability, changing the shape and/or
normalisation of the power spectrum. Even in the reflection model
there are differences in predicted variability from the more neutral
reflection seen in AGN compared to the much higher ionisation expected
for the hotter discs in BHB (Done \& Gardner 2011, in preparation).

The only clearcut distinction may be the high energy spectral shape
since this should be much less affected by atomic processes.  Low/hard
state spectra are intrinsically hard up to a thermal Comptonisation
rollover at a few hundred keV (e.g.  Ibragimov et al. 2005; Takahashi
et al. 2008; Makishima et al. 2008), while the high/soft and very high
states are soft and extend unbroken beyond 511~keV (Gierli\'{n}ski et al
1999; Zdziarski et al. 2001; Gierli\'{n}ski \& Done 2003).  We already know
that local AGN do show hard spectra in the 20-200~keV band, with a
clear high energy rollover (Zdziarski et al. 1995), but these
have mean luminosity below M1, so clearly correspond to a low/hard
state (Vasudevan et al. 2009).  Currently there are no objects in our
sample with well defined high energy spectra. Sensitive higher energy
data from NuSTAr or ASTRO-H on the objects in the M1/M2 sample should
give a clear test of their spectral state.

\section{Conclusions}

We explore the impact of the accretion disc temperature on the
appearance of an AGN. A standard disc model has a maximum maximum
temperature $T_{max}^4 \propto (L/L_{Edd})/M$ i.e. increases with
increasing $L/L_{Edd}$ and with decreasing black hole mass. However,
this temperature assumes that the radiation can completely thermalise,
which is not always the case. We follow the approach commonly used in
black hole binary systems and approximate results from full radiative
transfer calculations with a colour temperature corrected
blackbody. We show that this colour temperature correction is
substantial, with $f_{col}\sim 2.5$ for AGN with $T_{max}>10^5$~K
i.e. well above the temperature at which both H and He are fully
ionised so that electron scattering completely dominates the
opacity. The colour temperature correction is progressively smaller
for lower temperature AGN discs, becoming negligible for $T_{max}<
3\times 10^4$~K where H starts to become neutral so the associated
absorption edge opacity is large. The increase in colour temperature
correction with $T_{max}$ leads to an observed maximum disc
temperature which increases much faster than predicted from purely
blackbody models.  This increases the distinction in EUV and hence
emission line properties between objects with low $(L/L_{Edd})/M$
(typically Broad Line Seyfert 1 galaxies) and those with high
$(L/L_{Edd})/M$ (typical Narrow Line Seyfert 1s).

We show that the colour temperature correction is large enough that
the disc extends into the soft X-ray bandpass in the lowest
mass/highest mass accretion rate AGN such as the extreme NLS1
REJ1034+396. For these systems, much of the soft X-ray emission is
directly produced by the accretion disc unless the disc structure is
very different to that predicted. However, the shape of the 
disc emission does not match that of the 
observed 'soft X-ray excess' but requires Compton upscattering to 
be important. 

We assume that this Comptonisation takes place in the disc itself
rather than in a separate corona,  a scenario which is
supported by the lack of variability seen in this component 
in some extreme NLS1 (Middleton
et al. 2009; Jin et al. 2009). We build a self consistent model
for this, assuming that the disc structure changes inside some radius
$r_{corona}$, so the energy emitted within this emerges as Compton
upscattering rather than being able to thermalise to a (colour
temperature corrected) blackbody. 
We include the higher energy emission 
(power law tail) in our model by assuming
that part of the Compton upscattering takes place in a 
hot, optically thin corona
above and below the inner disc while the remainder is 
dissipated in the disc itself, but emerges as low
temperature/optically thick Comptonised emission rather than as a
blackbody. The key new aspect of our model is that the luminosity of
the soft excess and tail are constrained by energy conservation as we
assume that all the material accretes through the outer thin disc. 

The model fits well to the extreme NLS1 spectra, where there is a
'true' soft X-ray excess component seen above a weak X-ray tail which
has $\Gamma\sim 2-2.2$. The majority of the emission in these objects
is dissipated in the optically thick disc, with only $\sim 15$\% of
the bolometric flux required to form the soft X-ray excess, and $10$\%
producing the high energy tail. The small amount of energy in the soft
and hard Comptonisation components gives an implied $r_{corona}\sim
10-20~R_g$. At lower mass accretion rates we might expect that the
disc remains able to thermalise down to lower radii, so that the
spectra become even more disc dominated. However, the opposite is true
if the spectra are indicative of the intrinsic emission.  Another
issue with these spectra are that they appear rather similar to the
low/hard state in BHB, yet are seen up to $L/L_{Edd} \sim 0.2$, a
factor 10 higher than expected from BHB.  This would indicate some
process which breaks the mass scaling from BHB to AGN. 

Instead, multiple authors have suggested that the soft excess and
apparently hard 2-10~keV spectra arise as result of complex absorption
and/or reflection. If this is the case, the variability power spectra
will also be distorted, as clumpy absorption adds to the intrinsic
variability, as does the more neutral reflection expected in an AGN.
However, while the highest $L/L_{Edd}$ spectra do show complex, energy
dependent variability which may well indicate that these processes
shapes the spectrum, the lower $L/L_{Edd}$ objects, where the soft
excess component is most important in terms of the fraction of
bolometric luminosity, do not. The soft excess and hard spectra in
these objects appear instead to be intrinsic. Since these features are
not seen in BHB at these $L/L_{Edd}$ then this does require an
additional process to break the mass scaling from BHB to AGN. 

Extending the spectra beyond 10~keV
with {\it Suzaku}, {\it NuStar} and {Astro-H} may give the only
clearcut test of the correspondance between AGN and BHB states.

\section*{Acknowledgements}

We thank our referee, Bozena Czerny, for detailed comments and questions
which clarified our discussion of the colour temperature correction in AGN 
discs. CD acknowledges illuminating conversations on the
correspondance between AGN and BHB with Ian McHardy and Phil Uttley.

%-----------------------------------------------------

\appendix

\section{Comparison of {\sc optxagn} with {\sc dkbbfth}}

\begin{figure*}
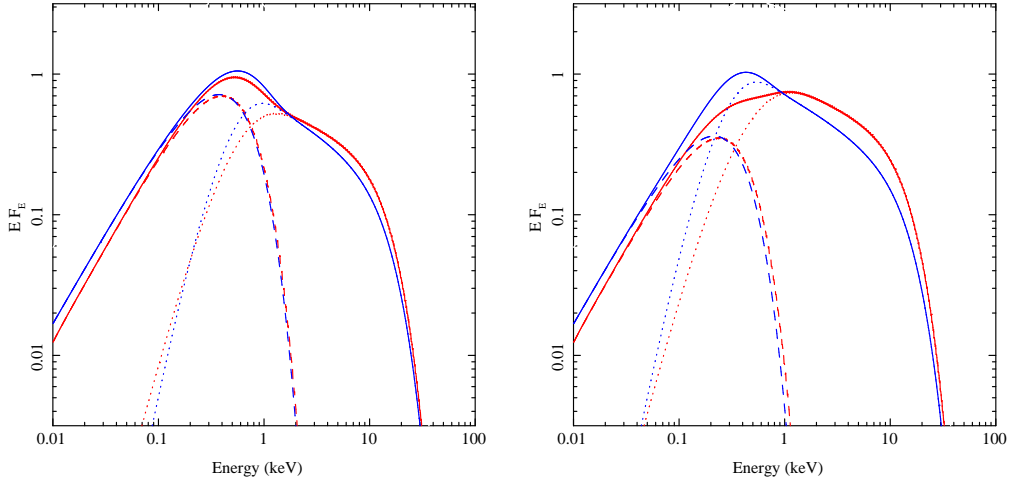

\centering
\begin{tabular}{cc}
\leavevmode  
\epsfxsize=6.5cm\epsfbox{dkbbfth_optxagn_m10_mdot18_r30_3.ps}
&
\epsfxsize=6.5cm \epsfbox{dkbbfth_optxagn_m10_mdot18_r75_3.ps}
\end{tabular}
\caption{Comparison of DK06 (red) and this model (blue) for a) 50\% and b)
  75\% of the total accretion power being Comptonised rather than
  radiated as a blackbody. The solid line shows the total model
  spectrum, which is made up of the outer disc (dashed line) and inner
  Comptonised emission (dotted line). The outer disc emission is
  always very similar, while the Comptonised emission can be 
  different due to the different assumptions about the seed photon
  energies. However, this only gives a significant difference where
  most of the accretion power is Comptonised (as in b). }
\label{fig:dk06}
\end{figure*}

Done \& Kubota (2006) (hereafter DK06)
developed such a similar model ({\sc dkbbfth}) in which the 
Comptonisation is energetically coupled to the accretion flow. 
This was applied to the 
very high state of Galactic Black Hole Binaries (BHB), where
the disc is strongly Comptonised by optically thick ($\tau\sim 2-3$)
plasma which has a lower temperature ($kT\sim 20$~keV) than the
typical temperatures of 100-200~keV seen at lower luminosity. DK06
assumed that the Comptonising corona formed a homogeneous plane
parallel atmosphere above the disc below some characteristic radius
$r_{corona}$, and that it is directly powered by accretion. Above this
radius the disc emitts all the energy dissipated by gravity as a
blackbody at the local temperature.  Below this, the energy is split
between the optically thick disc and corona.  More energy dissipation
in the corona implies less energy dissipated in the optically thick
disc, so its luminosity and temperature is lower. The local disc
photons act as the source of seed photons for the Comptonisation, so
the Comptonisation has to be done locally at many annulii in the
corona as this seed photon temperature varies with radius. This makes
the DK06 model extremely slow, so it is not feasible to use it to fit
a large sample of spectra.

Instead we develop a much faster model which keeps the key
aspect of energy conservation. We simply assume that the
energy within $r_{corona}$ is emitted as Compton upscattered
flux, with seed photons characterised by a blackbody at the disc 
temperature at $r_{corona}$. We also replace the underlying
Newtonian emissivity ($L(r)\propto r^{-3}$) used by DK06 with the
the fully relativistic Novikov-Thorne dissipation. 

We explicitly compare of our new model (blue) with that of DK06 (red)
in Fig. \ref{fig:dk06}. We assume a Schwarzschild black hole of
10\msun, accreting at $L/L_{Edd}=0.04$. This gives a maximum disc
temperature of $0.29$~keV, so we use this as the peak temperature of
the DK06 model and normalise the models so they have the same
flux. Fig. \ref{fig:dk06}a shows a comparison between the two models
for the case where 50\% of the accretion power is dissipated below
$r_{corona}$ (i.e. $12R_g$ for DK06, compared to $30R_g$ in the fully
relativistic emissivity), while Fig. \ref{fig:dk06}b is for 75\% of
the accretion power (i.e. $24R_g$ for DK06, compared to $75R_g$ in the
fully relativistic emissivity). The blackbody outer disc and
Comptonised inner disc components of each model are shown as a dashed
and dotted line, respectively.  The difference in radial emissivity
gives a slightly different shape to the disc emission for pure
blackbody spectra which is the reason for the slightly different
normalisation of the outer disc emission at low energies. As discussed
above, the two models are very similar except for the seed photon
energy for the Comptonised emission. This only makes a noticable
difference if most of the power is dissipated in the Comptonisation
region (i.e. $r_{corona}\ge 75~R_g$, and most of this difference is
confined to the unobservable EUV region. However, this does allow the
accretion energy to peak at a lower temperature than the equivalent
thermalised emission, which is probaly unphysical, so we urge caution
in interpretating the parameters when large $r_{corona}$ are derived
together with steep soft X-ray Comptonised spectra.

In all other circumstances this new model incorporates energy
conservation in the same way as {\sc dkbbfth} and gives results which
are very similar, yet is much faster so it is feasible to fit multiple
spectra.

\label{lastpage}

\end{document}